\documentstyle[12pt,epsf]{article}

\font\sf=cmss10                    %San-Serif 10
\def\half{\frac{1}{2}}
\def\ra{\rightarrow}

\def\ch#1#2{({#1 \atop #2 })}
\newcommand{\mathbold}[1]{\mbox{\boldmath $\bf#1$}}
\def\gcd{\mbox{gcd}}
\def\wE{\widehat{E}}
\def\mJ{\mathbold{J}}
\def\mA{\mathbold{A}}
\def\mB{\mathbold{B}}
\def\mC{\mathbold{C}}
\def\mX{\mathbold{X}}

\def\mz{\mathbold{z}}

\def\malpha{\mathbold{\alpha}}

\def\mdelta{\mathbold{\delta}}
\def\bbbz{{\sf Z\!\!\!Z}}
\def\sl2z{SL(2,\bbbz)}
\def\fracs#1#2{\textstyle\frac #1#2}

\def\det#1#2#3#4{\begin{array}{|cc|} #1 & #2 \\ #3 & #4 \end{array}\ }

\def\zz{{\bf z}}
\def\z0{{\bf z_0}}
\def\gg{{\cal G}}
\def\gh{\widehat{{\cal G}}}

\def\bbbp{\,{\hbox{P}\!\!\!\!\!\hbox{I}\,\,\,}}

\def\bbbc{{\hbox{C}\!\!\!\hbox{\rule{0.25mm}{2.88mm}\,\,\,}}}
\newcommand{\be}{\begin{equation}}
\newcommand{\ee}{\end{equation}}
\newcommand{\bea}{\begin{eqnarray}}
\newcommand{\eea}{\end{eqnarray}}
\newcommand{\nn}{\nonumber}
\newcommand{\bean}{\begin{eqnarray*}}
\newcommand{\eean}{\end{eqnarray*}}

\newcommand{\figref}[1]{Fig.~\protect\ref{#1}}

\newcommand{\mtimes}{\!\times\!}

\newcommand{\onefigure}[2]{\begin{figure}[htbp]
         \caption{#2\label{#1}(#1)}
         \end{figure}}
\renewcommand{\onefigure}[2]{\begin{figure}[t]
         \begin{center}\leavevmode\epsfbox{#1.eps}\end{center}
         \caption{#2\label{#1}}
         \end{figure}}
\begin{document}

\setlength{\baselineskip}{14pt}
\setlength{\parskip}{1.35ex}
\setlength{\parindent}{0em}
\renewcommand{\theequation}{\thesection.\arabic{equation}}

\noindent

\thispagestyle{empty}
{\flushright{\small MIT-CTP-2831\\hep-th/9902179\\}}

\vspace{.3in}
\begin{center}\Large {\bf Five-branes, Seven-branes and \\ Five-dimensional
$E_n$ field theories}
\end{center}

\vspace{.1in}
\begin{center}
{\large Oliver DeWolfe, Amihay Hanany, Amer Iqbal, Emanuel Katz}

\vspace{.1in}
{ {\it Center for Theoretical Physics,\\
Laboratory for Nuclear Science,\\
Department of Physics\\
Massachusetts Institute of Technology\\
Cambridge, Massachusetts 02139, U.S.A.}}
\vspace{.2in}

E-mail: {\tt odewolfe, hanany, amikatz@mit.edu, iqbal@ctpa04.mit.edu}
\end{center}
\begin{center}February 1999\end{center}

\vspace{0.1in}

\begin{abstract}
We generalize the $(p,q)$ 5-brane web construction of five-dimensional
field theories by introducing $(p,q)$ 7-branes, and apply this
construction to theories with a one-dimensional Coulomb branch.  The
7-branes render the exceptional global symmetry of these theories
manifest.  Additionally, 7-branes allow the construction of all $E_n$
theories up to $n=8$, previously not possible in 5-brane
configurations.  The exceptional global symmetry in the field theory
is a subalgebra of an affine symmetry on the 7-branes, which is
necessary for the existence of the system.  We explicitly determine
the quantum numbers of the BPS states of all $E_n$ theories using two
simple geometrical constraints.

\end{abstract}

\newpage

\section{Introduction}

Five-dimensional ${\cal N} = 1$ gauge theories have been much explored
in the last few years.  Those theories with a one-dimensional Coulomb
branch have a particularly rich variety of descriptions in string
theory.  They were first described by the dynamics of D4-branes in a
Type I$'$ background of D8-branes and orientifold O8-planes, where it
was determined that they possess enhanced exceptional global symmetry
at the origin of parameter space, which is a superconformal fixed
point \cite{seiberg}.  These theories also appear in M-Theory
compactifications on a Calabi-Yau threefold with a vanishing
four-cycle \cite{delpezzo,morrisonseiberg}.  Such complex surfaces,
called del Pezzo manifolds, are in one-to-one correspondence with the
five-dimensional theories with a single Coulomb modulus.  The BPS
spectrum can be obtained by using the intersection properties of
various curves in the del Pezzo \cite{kmv,mnw}.

Finally, some of the five-dimensional theories appear as the effective
field theory on a web of $(p,q)$ 5-branes in Type IIB \cite{ah, ahk,
kol,kolrahmfeld}.  The web has a single face, whose size corresponds to
the parameter of the Coulomb branch.  BPS states come from $(p,q)$
strings and string junctions stretched between the 5-branes.  This
realization is appealing, as parameters (such as masses and the gauge
coupling) and moduli of the theory may be easily read off the 5-brane
geometry.  However, it has its drawbacks.  The exceptional global
symmetry is not particularly evident, and must be inferred from the
correspondence with other presentations.  Moreover, this setup becomes
problematic beyond the $E_3$ configuration.  Those brane
configurations producing $E_4$ and $E_5$ involve parallel external
legs, which could result in six-dimensional modes.  Those
corresponding to $E_6$, $E_7$ and $E_8$ are worse, for the external
legs would cross, and the configurations cannot exist.  This is
related to the lack of a toric description for the corresponding del
Pezzos \cite{leungvafa}.
  
In this paper we generalize the $(p,q)$ 5-brane setup so as to resolve
these problems.  We do this by introducing $(p,q)$ 7-branes to the
picture.  This can be done at no additional cost in supersymmetry.
The external $(p,q)$ 5-branes are allowed to terminate at finite
distance on the appropriate $(p,q)$ 7-brane.  Motion of each 7-brane
along the direction of the 5-brane does not affect the
five-dimensional theory.  One may then bring the 7-branes within the
5-brane face, so that the external legs vanish altogether.  The BPS
states of the theory are hypothesized to switch from being 5-5 string
junctions to 5-7 junctions for certain regions of parameter space.

{From} this point of view the global symmetries are manifest, as the
technology of enhanced 7-brane algebras can be brought to bear
\cite{gz, mgthbz, dewolfezwiebach, dhiz, dewolfe, finite, infinite}.
The symmetry algebras that appear on 7-brane configurations have been
classified \cite{finite,infinite}, and it is possible to determine the
weight vector associated to any string junction with support on such a
configuration \cite{dewolfezwiebach}.  Moreover, the string junctions
have a natural intersection form inherited from curves in K3
\cite{dhiz, nekrasov}, which may be used to determine the BPS spectrum
in a fashion similar to the del Pezzo case.  In fact, the symmetries
of the 7-branes turn out to be not just $E_n$, but the affine algebras
$\wE_n$, first observed in \cite{dewolfe}.  We will show how the
constraint that a junction end on the 5-branes reduces the effective
symmetry to just $E_n$, consistent with our expectations for the field
theory.  The affine character of the 7-branes is not irrelevant,
however; the 5-branes themselves have an associated weight vector
corresponding to the imaginary root of the affine algebra, and in fact
a 7-brane configuration must be affine in order to support a 5-brane
web with a single face.

Moreover, with both 5-branes and 7-branes we can describe any of the
$E_n$ theories, all the way up to $E_8$.  This is because there is no
obstruction to placing certain additional 7-branes inside the 5-brane
web, thereby forming the entire $E_n$ series.  Then for $n \geq 4$,
these 7-branes simply cannot be brought out of the 5-brane web and
moved off to infinite distance without creating the problem of
parallel/crossing legs.  When the face shrinks to zero size, we find
that for $n \leq 3$ the configuration reverts to having no 7-branes
within the face before reaching the superconformal point.  For
precisely the $n \geq 4$ cases where no configuration existed without
7-branes, the fixed point occurs when the face collapses around a
subset of the 7-branes.  The finite algebra $E_n \subset \wE_n$ is
restored on this subset of branes.

In the section 2 we will recapitulate the $(p,q)$ 5-brane web setup
and the nature of the five-dimensional gauge theories.  Then in
section 3 we discuss placing 7-branes in the background and allowing
5-brane external legs to end on them, and section 4 describes how when
the 7-branes are moved inside the face, the 5-5 strings can be
replaced by 5-7 strings.  Section 5 reviews the way string junctions
realize algebras on 7-branes, and shows how the $E_n$ theories
correspond to $\wE_n$ brane configurations.  The fact that only a
finite symmetry is realized on the spectrum is explained.  In section
6 we detail the case of $E_1$ explicitly.  We then compare 5-7-brane
techniques to the del Pezzo realization of five-dimensional theories
and explore the predictions for BPS spectra in section 7.  Finally
section 8 discusses the approach to the superconformal point.  We
conclude with some speculative remarks.

\section{Five-dimensional theories and (p,q) 5-brane webs}
\label{review}

We are concerning ourselves with the study of five-dimensional gauge
theories with 8 supercharges.  The vector multiplets contain one real
scalar, and we are interested in those theories with a one-dimensional
Coulomb branch associated to a single vector multiplet.  We denote the
various theories by their associated exceptional global symmetry,
$E_n$.  For a discussion of more general five-dimensional
theories with 8 supercharges, see \cite{ims,telaviv}.

Most of these theories, the ones with $1 \leq n \leq 8$, can be
realized as the low-energy limit of $SU(2)$ with $N_f = n-1$
hypermultiplets.  The exceptional symmetries were found by Seiberg
\cite{seiberg} using Type I$'$ arguments.  The restriction $N_f \leq
7$ was motivated from field theory as well.  Two additional theories,
$\tilde{E}_1$ and $E_0$, were identified by flowing from the $E_2$
theory \cite{morrisonseiberg}. The former has a $u(1)$ global
symmetry, and is identified as the IR limit of a pure $SU(2)$ gauge
theory with nontrivial $\bbbz_2$ $\theta$-angle \cite{delpezzo}, while
the latter has no global symmetry and no interpretation as a gauge
theory since there are no W-bosons on the Coulomb branch.

Five-dimensional gauge theories are characterized by a gauge coupling
$1/g^2$ with the dimensions of mass, as well as (real) masses for
possible hypermultiplet matter.  The prepotential is restricted to be
cubic.  Even in the absence of a cubic term classically, one is
generically generated at one-loop.  A classical cubic term is an
additional parameter, and can be thought of as having arisen from
matter already integrated out.  The BPS spectrum includes not just
W-bosons and quarks, but also instantons, which are particles in 5D.
Furthermore, the BPS spectrum also features magnetic strings.

Let us review the $(p,q)$ 5-brane setup.  The $(p,q)$ 5-branes fill
the common directions 01234, which is the spacetime of the effective
five-dimensional theory.  In addition each 5-brane occupies
1-dimensional support in the 56 plane, with the slope of the curve
determined at each point by the charges $(p,q)$ of the 5-brane and the
expectation value of the complexified coupling $\tau = \chi + i e^{-
\phi}$:
\begin{eqnarray}
\label{slope}
\Delta x_5 + i \Delta x_6 \, || \, p + \tau q \,.
\end{eqnarray}
The 5-branes occupy a point in the 789 directions, which will not play
a role in what follows.  Three 5-branes can meet at a point as
long as charge is conserved: $\sum_i (p_i, q_i) = 0$.  Thus we obtain
a $(p,q)$ 5-brane web.

The slope condition stems from the requirement of supersymmetry.
(Together with charge conservation, it can also be thought of as the
condition for mechanical equilibrium.)  Type IIB string theory has 32
supercharges, organized into two 16-component Majorana-Weyl spinors
$Q_L$, $Q_R$ with the same chirality: $\overline{\Gamma} Q_L = Q_L$,
$\overline{\Gamma} Q_R = Q_R$, where $\overline{\Gamma} = \Gamma_0
\cdots \Gamma_9$.  If we place a D5-brane (= (1,0)-brane) in the
012345 directions, the supersymmetries preserved are $\epsilon^L Q_L +
\epsilon^R Q_R$ with the constant spinors $\epsilon$ satisfying
\begin{eqnarray}
\label{D5susy}
\epsilon^L = \Gamma_0 \Gamma_1 \Gamma_2 \Gamma_3 \Gamma_4 \Gamma_5
\epsilon^R \,,
\end{eqnarray}
which as usual preserves $1/2$ of the supersymmetry of the vacuum.  If
in addition we place an NS 5-brane (= $(0,1)$-brane) in the 012346
directions, we have the additional constraints
\begin{eqnarray}
\label{NS5susy}
\epsilon^L = \Gamma_0 \Gamma_1 \Gamma_2 \Gamma_3 \Gamma_4 \Gamma_6
\epsilon^L \,, \quad \quad \epsilon^R = - \Gamma_0 \Gamma_1 \Gamma_2
\Gamma_3 \Gamma_4 \Gamma_6 \epsilon^R \,,
\end{eqnarray}
and a total of $1/4$ of the supersymmetry is preserved.  If any
$(p,q)$ 5-brane is added to the configuration, no additional
supersymmetry will be broken as long as it fills the 01234 directions
and a line in the 56 plane determined by (\ref{slope}).
Thus the $(p,q)$ 5-brane web has an effective five-dimensional field
theory with 8 supercharges, or ${\cal N} = 1$.  Each face of the
5-brane web corresponds to a modulus on the Coulomb branch, and thus
we will be considering configurations with one face.  

\onefigure{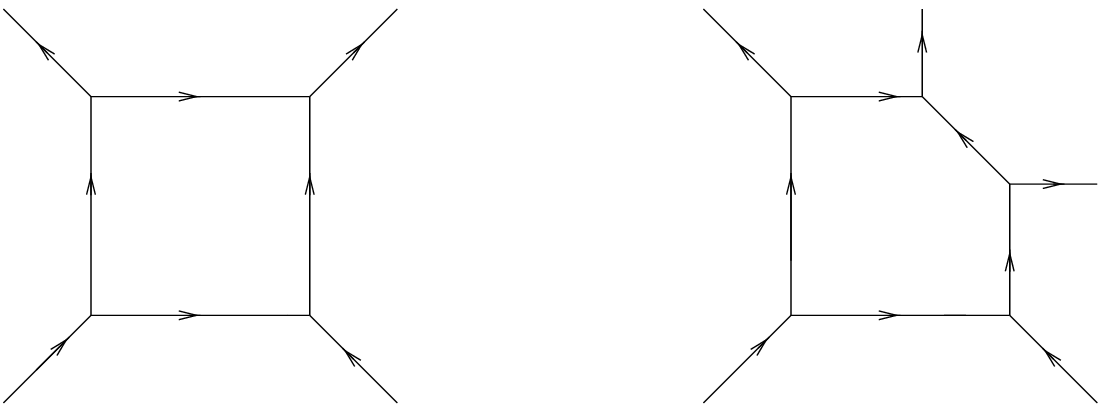}{a.\ The 5-brane web corresponding to the $E_1$
theory. b.\ The 5-brane web corresponding to the $E_2$ theory, with
horizontal external leg (D5-brane) generating a flavor.}
The BPS particles in the spectrum arise from strings and string
junctions stretching between the 5-branes, while the BPS strings are
3-branes wrapping the face.  IIB strings are also characterized by
charges $(p,q)$, and they form junctions and webs in the plane as
5-branes do.  A $(p,q)$ string can end on a $(p,q)$ 5-brane, and they
must be perpendicular at the contact point; thus the slopes of string
segments are rotated 90 degrees from those of 5-branes (\ref{slope}).

In \figref{5-d-3}a, we show the 5-brane web for the $E_1$ theory, with
the value $\tau = i$ chosen to make the slopes of the 5-branes agree
with their charge vectors; in the absence of 7-branes $\tau$ is a
constant, and its value is a superfluous parameter from the point of
view of the five-dimensional theory \cite{ah,ahk}.  The face is made
of two D5-branes and two NS5-branes; the external legs have charges
$(1,1)$ and $(1,-1)$.  The basic BPS states are the W-boson, realized
as a F-string connecting the D5-branes, and the instanton, a D-string
stretching between NS5-branes; these are a doublet of the $E_1 =
SU(2)$ global symmetry, and other BPS states can be thought of as
linear combinations of these.  This field theory is pure $SU(2)$ gauge
theory (with vanishing $\theta$-angle); flavors are associated with
D5-brane external legs, as in the realization of $E_2$ in
\figref{5-d-3}b.

Let us demonstrate that 7-branes may be added to the configuration
without breaking any more supersymmetry.  Combining (\ref{D5susy}) and
(\ref{NS5susy}), we obtain
\begin{eqnarray}
\epsilon^L &=& - (\Gamma_0 \Gamma_1 \Gamma_2 \Gamma_3 \Gamma_4
\Gamma_5) (\Gamma_0 \Gamma_1 \Gamma_2 \Gamma_3 \Gamma_4 \Gamma_6)
\epsilon^R \,, \\ &=& - \Gamma_5 \Gamma_6 \epsilon^R \,, \nonumber
\end{eqnarray}
but since $\overline{\Gamma} Q_{L,R} = Q_{L,R}$ and
$\overline{\Gamma}^2 = 1$, we have $\overline{\Gamma} \epsilon^{L,R} =
\epsilon^{L,R}$ as well, and then
\begin{eqnarray}
\epsilon^L &=& - \Gamma_5 \Gamma_6 \overline{\Gamma} \epsilon^R \,, \\
           &=& \Gamma_0 \Gamma_1 \Gamma_2 \Gamma_3 \Gamma_4 \Gamma_7
           \Gamma_8 \Gamma_9 \epsilon^R \,,
\end{eqnarray}
which is recognized as the constraint on the preserved supersymmetries
in the presence of a D7-brane at a point in the 56 plane and filling
the other directions.  In fact, all $(p,q)$ 7-branes preserve this
same supersymmetry.  Thus we can introduce 7-branes to the 5-brane web
without breaking any additional supersymmetry.  This is reminiscent
of the introduction of D3-branes to a different D5/NS5 system with a
three-dimensional effective field theory \cite{hananywitten}.  Since
7-branes are a source for $\tau$ the condition (\ref{slope}) will
become a complicated function of position in the 56 plane.

\section{7-branes and 5-brane webs}

\subsection{Ending the 5-brane web on 7-branes}

\onefigure{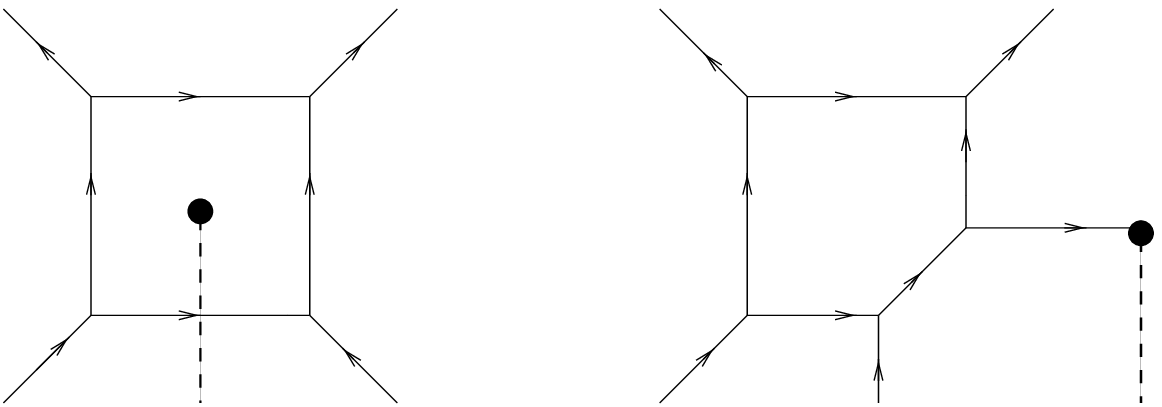}{A D7-brane added to the $E_1$ configuration, adding
a flavor and producing $E_2$.}  Consider the $E_1$ web configuration
with a D7-brane inserted; we have placed the 7-brane branch cut where
it will not affect the 5-brane system, as shown in \figref{5-d-9}a.
Fundamental strings can stretch between the D7-brane and the
D5-branes, creating quark hypermultiplets in the fundamental
representation of SU(2), the gauge group of the 5D theory.  The length
of these strings corresponds to the mass of the quarks, and thus the
vertical position of the D7 is the quark bare mass parameter.  Moving
the D7 brane horizontally within the face does not change the length
of the strings, and therefore does not affect any parameters of the
theory \cite{hananywitten}.

In fact, one can even move the D7 brane horizontally past the NS5
brane.  Via the familiar brane creation \cite{hananywitten}, a
D5-brane prong will form, going to the right of the NS5-brane and
ending on the D7-brane, as in \figref{5-d-9}b.  The quarks now result
from strings going between the original D5-branes in the face and the
newly-created D5, and as we move the D7-brane horizontally, the quark
length and hence the mass parameters do not change.  Thus horizontal
motion of the D7 is irrelevant to the field theory everywhere.  Since
we can move the D7-brane to infinity, we conclude that from the point
of view of the 5D theory a semi-infinite D5-brane is equivalent to a
finite D5-brane ending on a D7.  This is consistent with the fact that
both the D7 brane and the semi-infinite D5 brane are infinitely massive
from the 5D perspective.  This makes their vertical motion a parameter
of the theory.

One can tune the locations of the external legs such that as the face
shrinks, the NS5-branes collapse together before the D5-branes do.  In
this case the situation can be interpreted as an $SU(2)$ living on the
NS5-branes, rather than the D5-branes, thanks to S-duality; this is
known as continuation past infinite coupling \cite{ah,ahk}.  Quarks in
the NS5 theory can be provided by external NS5 legs, which we can
analogously end on $(0,1)$ 7-branes, again expecting that motion of
the 7-brane along the path of the NS5 will not affect the parameters
of the 5D theory; these parameters are related to the parameters and
moduli of the theory before continuation past infinite coupling.  Thus
in general, we may take both NS5 and D5-branes to end on $(0,1)$ and
$(0,1)$ 7-branes respectively, and motion along the external legs
should not affect any parameters or moduli of the D5 brane theory.

\onefigure{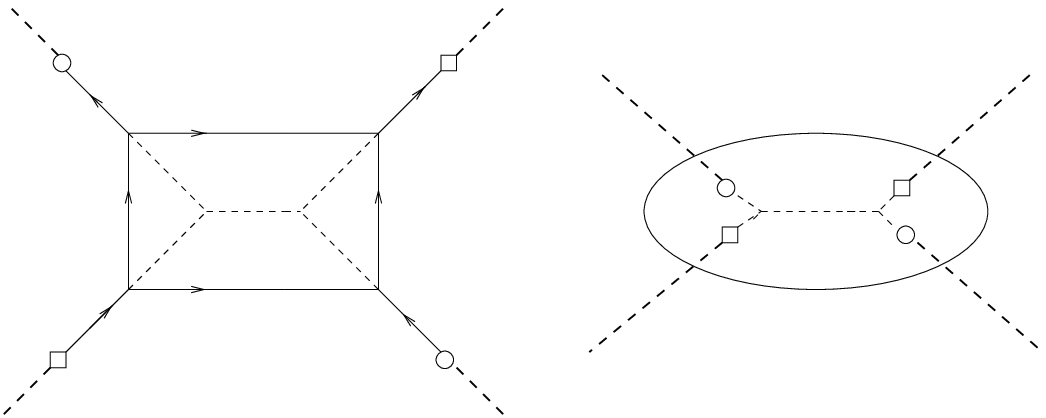}{a.\ The external 5-branes of the $E_1$ web can end
on 7-branes. b.\ The 7-branes move within the 5-brane face, removing
the external 5-branes entirely.}  Since this is true for both types of
7-branes, we propose that any semi-infinite $(p,q)$ 5-brane leaving
the $(p,q)$ web may be replaced with a finite 5-brane ending on the
appropriate $(p,q)$ 7-brane.  This 7-brane can be moved along its
prong without changing the 5D field theory.  For example, the pure
$SU(2)$ theory configuration can be modified so that all external legs
end on 7-branes, as in \figref{5-d-4}a.  The dashed lines represent
the 7-brane branch cuts, which point away from the web so as not to
disturb the charges of the 5-branes.  The dotted lines show the web at
the origin of the Coulomb branch.  Note that the presence of 7-branes
creates a non-trivial metric.  Thus the 5-branes follow curved
geodesics, rather than straight lines of a given slope, and we will
use the slopes merely to indicate the 5-brane charges.

Let us now ask what would happen if we move all the 7-branes, along
their prong directions, into the center of the web.  As each brane
passes the corner to which it is connected, brane creation will occur
and the 5-brane prong will disappear.  The result will be a single
5-brane loop going around the 7-branes, with its $(p,q)$ nature
changing as it passes the various branch cuts, as in \figref{5-d-4}b.
Since the loop is closed, this is indeed a five-dimensional theory.

Evidence for the existence of such a loop may be found by taking the
parameter $1/g^2$ to be very large, resulting in a short and wide web,
and then taking the size of the face sufficiently large.  The four
7-branes become two $(1,-1)$ and $(1,1)$ pairs, each of which is the
quantum resolution of the O7 orientifold \cite{sen}.  For sufficiently
large modulus they can be treated as two O7-planes, and in this limit
the geodesic corresponding to the loop may be explicitly constructed.
Indeed, its length is found to be independent of the size of the face,
as expected from a modulus deformation of the web.  In addition,
performing a T-duality transformation along the 5-brane direction, one
obtains Seiberg's IIA setup of the same gauge theory.  Here, the D5
brane becomes a D4 brane, and the two O7-planes merge into a single
O8-plane.

\subsection{BPS states as 5-7 strings}

In the case of the familiar transition of a pronged D7 brane across 
an NS5 brane, it is known that certain states of the 5D theory 
(namely the quarks) which came from D5-D5 strings, instead originate
from D5-D7 strings.  We would like to show that upon
moving the 7-branes into the face of the $SU(2)$ web, every BPS state
described by 5-5 strings may be alternatively described by 5-7 strings.
Here, a description entails understanding of a state's quantum numbers
under the local and global symmetries.  Since the states we consider are
BPS, this information is also enough to proscribe their masses.  

\onefigure{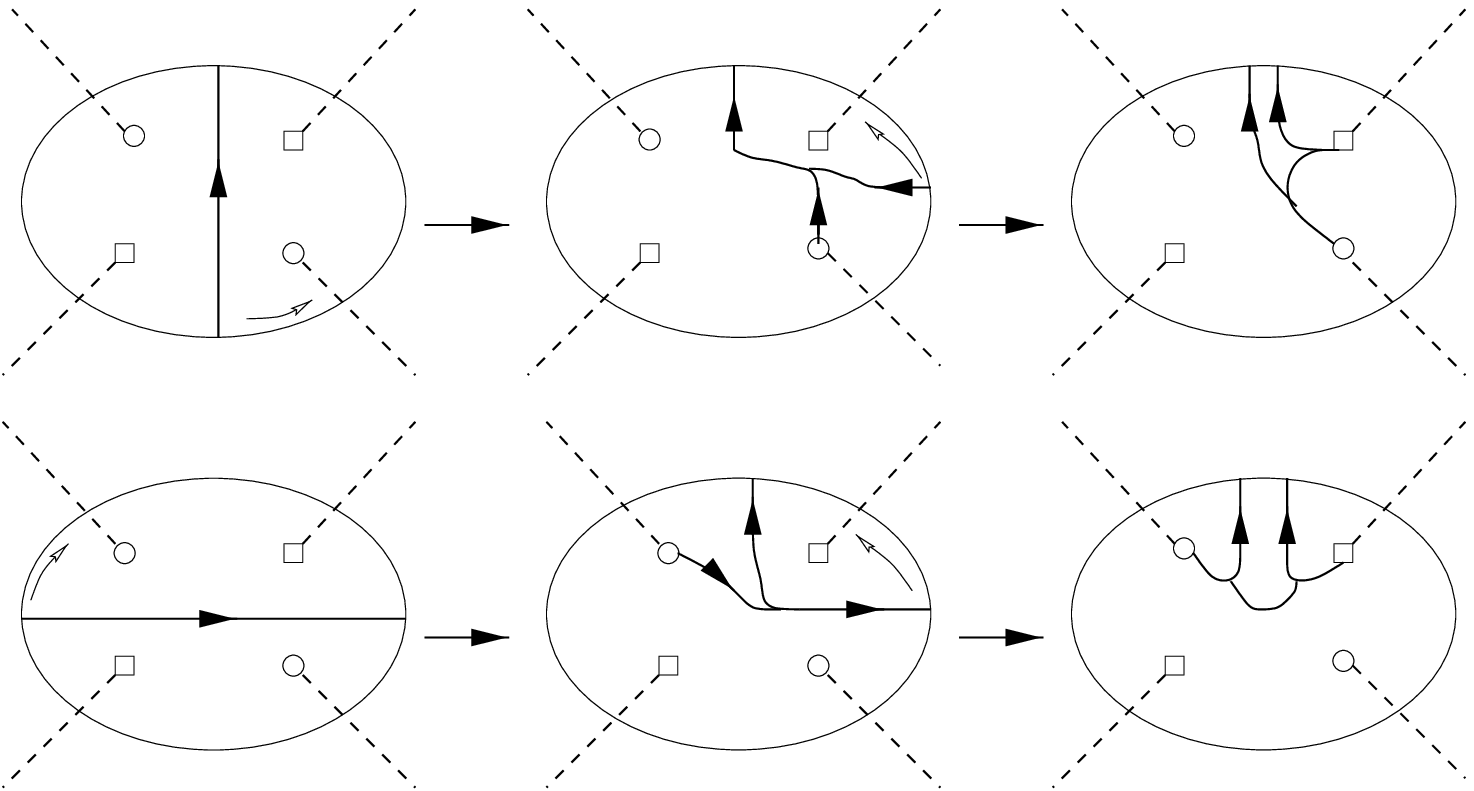}{The W-boson (top) and instanton (bottom) can be
continuously deformed from being 5-5 stings to 5-7 string junctions.}
The map between 5-5 and 5-7 strings is achieved through the continuous
deformation of the 5-5 string junctions.  Once the 7-branes are inside
the face, the web consists of a single 5-brane loop.  Hence, one may
drag the endpoints of each 5-string junction until they lie, for
example, entirely on the top (D5-brane portion) of the loop.  As the
junction is deformed in such fashion, prongs on 7-branes will form via
brane creation.  The result is a 5-7 string junction, starting at the
7-branes and ending on the top of the 5-brane loop.  For the cases of
the instanton and the W-boson this process is displayed in
\figref{5-d-5}.

Let us stress that metric around the 7-branes is nontrivial, and
therefore for generic locations of the 7-branes we do not know if the
actual BPS (geodesic) junction is a 5-5 or a 5-7 string.  Rather,
%because the junction is BPS,
we do not expect these deformations to change the quantum numbers
associated to the BPS state.
One way to see this is to note that Type IIB-theory with 7-branes on a
sphere is dual to M-theory on a K3-surface, and string junctions
correspond to membranes wrapping complex curves in K3.  The quantum
numbers of a junction are determined by the homology cycle wrapped by
the membrane and its intersection numbers, which are invariant under continuous transformations.  Hence, to describe the junction curve we may
choose any convenient homology representative, whether it is the BPS
(area minimizing) one or not.

%The various
%quantum numbers of the junction curve follow from its topology, and
%from its intersection numbers with other curves.  However, continuous
%deformations of a curve do not affect its topology nor its
%intersection numbers.  

The advantage of representing the states of the theory as 5-7 string
junctions is that the action of the global symmetry becomes more
transparent.  Each 7-brane has a $U(1)$ gauge theory living on its
worldvolume. When 7-branes of various types are able to become
coincident, gauge symmetry enhancement will occur in their worldvolume
theory.  The string junctions leaving the 7-branes will thus fall
naturally under representations of this enhanced symmetry group.
Therefore, this enhancement translates into a global symmetry in the
5D theory.  This is the subject of the next section.

\section{String junctions and 7-branes}

We now turn to a review of the symmetries arising on $(p,q)$ 7-brane
configurations, and how they are realized on $(p,q)$ strings ending
on the 7-branes.  Using the notation of \cite{gz}, we will often
refer to 7-branes with charges $[1,0]$, $[1,-1]$ and $[1,1]$ as $\mA$,
$\mB$ and $\mC$-branes, respectively.

A configuration of $(p,q)$ 7-branes generates an eight-dimensional
worldvolume gauge theory, whose gauge group $\gg$ is determined by the
charges of the 7-branes when their branch cuts are given a
well-defined ordering, generalizing the familiar $U(n)$ gauge group
that arises when $n$ 7-branes with the same $(p,q)$ charge coincide.
In addition, $\gg$ appears as a global symmetry on the worldvolume of
some probe brane or branes brought near the 7-brane configuration.
Thus in our example, we expect the 7-branes to induce a global
symmetry in the five-dimensional theory.

The objects charged under the symmetry of the 7-branes are various
strings and string junctions; those beginning and ending on the
7-branes are the adjoint representation, while various other
representations appear as string junctions beginning on the 7-branes
and ending on another object, such as a 5-brane.  Only a $(p,q)$
string can end on a $(p,q)$ 7-brane.  String junctions stretch in the
plane orthogonal to the 7-branes, and may end on one another in
agreement with charge conservation to form webs just as the 5-branes
do.  The slopes are a ninety-degree rotation from those of the
5-branes, as discussed in section \ref{review}.

A given string junction $\mJ$ with support on $n$ 7-branes is
characterized by $n$ invariant charges $Q^i$, where $i$ labels the
7-branes.  The $Q^i$ are integers, and are a combination of the number
of string prongs ending on a given brane and the number of string
segments passing that brane's branch cut.  They measure how much
charge flows out of each 7-brane into $\mJ$, and are invariant under
brane creation transformations.

The space of junctions is a lattice, and it has an inner product
defined on it, the intersection number. This quantity can be thought
of as being inherited from the intersection of curves in K3.  (For
more details, see \cite{dewolfezwiebach}.)  As discussed in
\cite{dewolfezwiebach}, the intersection of two junctions $\mJ$ and
$\mJ'$ receives contributions from each brane on which both have
prongs, and from points where a string segment with charges $(p,q)$
belonging to one junction crosses a segment with charges $(r,s)$
belonging to the other:
\begin{eqnarray}
\label{intersection1}
(\mJ, \mJ') = - \sum_{i=1}^n n_i n_i' + \sum_{\mbox{crossings}}
\det{p}{r}{q}{s} \,,
\end{eqnarray}
where $n_i$ is the number of prongs leaving the $i$th brane minus the
number entering it, and the $(p,q)$ and $(r,s)$ segments are ordered
in a counterclockwise fashion.  This can be expressed entirely in
terms of the $Q^i$ and $Q'^i$, but we shall not do so here.

Alternately, a junction can be specified by its Lie algebra weight
vector $\lambda$ with respect to the group $\gg$, which includes
possible $u(1)$ charges and is given by $n-2$ Dynkin labels, and the
asymptotic charges carried away from the 7-branes, $p$ and $q$.  In
terms of these quantities the intersection form is just
\begin{eqnarray}
\label{intersection}
(\mJ, \mJ') = -\lambda \cdot \lambda' + f(p,q;p',q') \,,
\end{eqnarray}
where $\lambda \cdot \lambda'$ is exactly the usual Lie algebraic
inner product on weight vectors, and $f$ is a certain expression
quadratic in the charges and symmetric between $(p,q)$ and $(p',q')$,
which is specified by the total monodromy of the 7-branes
\cite{finite}.  For the root junctions the asymptotic charges vanish,
and the intersection of the simple root junctions is just minus the
Cartan matrix of $\gg$.

A necessary condition for a junction to be BPS is that it obey
\cite{dhiz}
\begin{eqnarray}
\label{jsquared}
(\mJ, \mJ) \geq -2 + \gcd(p,q) \,.
\end{eqnarray}
This is because BPS junctions are associated to holomorphic curves in
K3, and hence must satisfy $(\mJ, \mJ) = 2g + b -2$ where $g \geq 0$
is the genus and $b = \gcd(p,q)$ is the number of boundaries of the
curve.

7-brane configurations realizing the simply laced algebras $A_n$,
$D_{n \geq 4}$ and $E_6$, $E_7$, $E_8$ correspond to Kodaira
singularities of K3 when the branes are brought to a point.  Other
brane configurations cannot be so collapsed, such as the $D_{n \leq
3}$ associated to Seiberg-Witten probe theories and $E_{n \leq 5}$.
There are distinct 7-brane configurations for $E_1$ and $\tilde{E}_1$
with $SU(2)$ and $U(1)$ symmetry, respectively, as well as $E_0$ with
no symmetry, exactly as in the 5-brane case.

Perhaps surprisingly, there also exist 7-brane configurations whose
associated string junctions fill out the adjoint representation of an
infinite-dimensional algebra.  These cases are never collapsible, so
the algebra should not be thought of as being a familiar gauge
symmetry.  In \cite{dewolfe}, it was shown how the affine versions of
all the exceptional algebras, $\wE_n$, appear when a certain new
7-brane is added to the regular $E_n$ groupings.  Interestingly, these
are the only affine algebras realized on 7-branes \cite{infinite}.

Returning to the problem of the 7-branes in the 5-brane web, we now
ask the question of what symmetry is associated to the 7-branes with
the same $(p,q)$ charges as the external 5-brane legs they replaced.

The answer is that for the $E_n$ 5D theory, the full global symmetry
of the 7-branes turns out to be $\wE_n$.  Let us return to the example
of the $E_1$ theory.  The four associated 7-branes form the
configuration $\mB \mC \mB \mC$.  There are two obvious root
junctions, the BB and CC strings, which we will call $\malpha_0$ and
$\malpha_1$. They have self-intersection $(\malpha_i, \malpha_i) =
-2$, and intersection $(\malpha_0, \malpha_1) = 2$, and thus are
simple roots for an $\widehat{su(2)} = \wE_1$ algebra.  

For the other cases where 5-brane webs realize an $E_n$ theory, the
associated 7-branes also have $\wE_n$ symmetry, even for the cases
$E_4$ and $E_5$ where there were parallel external legs and the
five-dimensional theory was ill-defined.  As we shall see, the
sequence of 7-branes may be continued all the way up to an $\wE_8$
configuration, which will realize the $E_8$ five-dimensional theory.

One might be tempted to conclude from this that the 5D theories should
realize an infinite-dimensional global symmetry.  However, this proves
not to be the case.  Before explaining why this is so, let us briefly
review a few properties of affine algebras.

Each finite simple compact Lie algebra $\gg$ has an affine extension
$\gh$.  If $T^a$ generate $\gg$, $\gh$ has the generators $\{T^a_n, K,
D\}$ where the grade $m \in \bbbz$, and the commutation relations are
\begin{eqnarray}
\left[ T^a_{m_1} , T^b_{m_2} \right] &=& f^{ab}_{~\;\;c} \,
T^c_{m_1+m_2} + \kappa^{ab} \, m_1 \, \delta_{m_1+m_2} K \,, \nn \\
\left[ D, T^a_m \right] &=& - m \, T^a_m \,, \\
\left [K, T^a_m \right] &=& \left[ K, D \right] = 0 \,, \nn
\end{eqnarray}
where $f^{ab}_{~\;\;c}$ and $\kappa^{ab}$ are the structure constants
and Killing form of $\gg$, respectively.

Let $\{ H^i, E^{\alpha} \}$ be a set of Cartan and root generators for
$\gg$.  The Cartan subalgebra of $\gh$ is $\{ H^i_0, K, D \}$.  In
addition to an infinite number of roots associated to the generators
$E^{\alpha}_m$, there are also roots associated to $H^i_m$, $m \neq
0$.  These are all integer multiples of the imaginary root $\delta$.
The imaginary root has a few interesting properties, including $\delta
\cdot \alpha_i =0$ for any root, including $\delta$ itself.

Affine algebras have highest weight representations as finite algebras
do, but differ in having no lowest weight.  The level $k(\lambda)
\equiv \delta \cdot \lambda$ is a constant over a representation, as
subtracting simple roots will not change it.

When affine algebras are realized on junctions, the level turns out to
be a linear combination of $p$ and $q$.  (Affine configurations are in
fact the only ones where the asymptotic charges correspond to a Lie
algebraic quantity.)  The level can be determined by intersection with
the imaginary root junction $\mdelta$, $k(\mJ) = - (\mJ, \mdelta)$.
The imaginary root junction obeys $(\mdelta, \mdelta) = 0$, and since
for a general junction with no asymptotic charges $\mJ^2 = 2g-2$, we
see that unlike the usual root junctions with $g=0$, $\mdelta$ is a
genus one object.  As such it can be realized as a closed loop
surrounding the 7-branes.  The $(p,q)$ charge of a segment of the loop
changes as it passes each branch cut, but after passing all the branch
cuts it comes back to itself, a situation only possible when the
7-branes realize an affine algebra.  The intersection is
\begin{eqnarray}
\label{affineinter}
(\mJ, \mJ') = - \lambda \cdot \lambda' + f(p,q;p',q') - m k' - m' k \,,
\end{eqnarray}
where $\lambda \cdot \lambda'$ and $f$ are the same as for the
configuration realizing the corresponding finite algebra, and $m$ is
the grade of the junction, given by the number of $\mdelta$ factors it
contains.  For the $E_n$ case we have
\begin{eqnarray}
\label{f}
f(p,q;p',q') = \frac{1}{9-n} \left( pp' - \fracs12 (n-3) (pq' + qp') +
(2n-9) qq' \right) \,.
\end{eqnarray}

Let us present $\mdelta$ for the $\mB \mC \mB \mC$ configuration,
which is $\wE_1$. A $(1,0)$ string may cross the first $\mB$ branch
cut, becoming a $(0,1)$ string, and then cross the first $\mC$ cut,
becoming $(-1,0)$.  The second pair of cuts turn it into $(0,-1)$ and
then $(1,0)$ again, after which it is free to loop around and join
itself. Higher $\wE_n$ configurations are obtained by adding
$\mA$-branes, whose branch cuts do not affect the $(1,0)$ string, so
$\mdelta$ exists as before.  $\widehat{\tilde{E}}_1$ and $E_0$ are
slightly different, but $\mdelta$ still starts and ends as a $(1,0)$
string; details can be found in \cite{dewolfe, infinite}.

Now consider the 5-brane web configuration.  Once a 7-brane moves
inside the face, the external leg vanishes, and instead it is the
7-brane branch cut that changes the $(p,q)$ character of the
5-brane. When all 7-branes are inside the face, the entire 5-brane
configuration now forms a loop around them, changing $(p,q)$ labels at
each branch cut but coming back to itself after all branch cuts.  In
fact, the 5-brane web in each case traces the same curve with the same
charges as the imaginary root junction $\mdelta$.

This is an interesting result.  Recall that only affine 7-brane
configurations can support nontrivial loops with one face; the affine
character is thus essential for the existence of the 5-brane web.
This provides an oblique reason why only exceptional affine algebras
are realized by 7-brane configurations; others would lead to new
five-dimensional theories with other global symmetries, which are not
seen in the other string realizations of these theories.  In
particular, the affine exceptional 7-brane configurations are in
one-to-one correspondence with del Pezzo surfaces.

We have suggested that there should be regions in parameter space
where the BPS states are 5-7 strings; this turns out to be a useful
way to obtain the quantum numbers of states, since the 7-branes make
the global symmetry manifest.  However, we must explain why the
affine character of the 7-branes is invisible to the field theory.

A junction leaving an $\wE_n$ configuration of 7-branes is
characterized by its finite $E_n$ weight vector $\lambda$, a grade
$m$, and charges $p$ and $q$; one linear combination of these charges
determines the level $k$.  However, there is a restriction on the
possible quantum numbers of a junction that appears in the 5D theory:
its $(p,q)$ charges must match those of the segment of 5-brane on
which it ends.  But since the 5-brane loop traces the same curve as
$\mdelta$, the level of a junction $\mJ$ that can end on the 5-brane
is, from (\ref{intersection1}):
\begin{eqnarray}
k(\mJ) \equiv - (\mdelta, \mJ) &=& \det{p}{p}{q}{q} \,, \\
&=& 0 \,. \nn
\end{eqnarray}

Thus the restriction on junctions which can end on the 5-brane
configuration is precisely that they have vanishing level.  This will
be the reason why only a finite remnant of the affine algebra survives
as a global symmetry in the five-dimensional spacetime.

For definiteness, let us consider junctions that end on the top
$(1,0)$ 5-brane; as before, junctions ending elsewhere on the 5-brane
loop can be slid along until they end at the top, possibly crossing
branch cuts or undergoing brane creation along the way.  To end on
this 5-brane the junctions must satisfy $q = 0$, and in fact in this
presentation $k = -q$ \cite{dewolfe}.  Now $p$ is related to the
electric charge of the state under the Cartan generator of the 5D
$SU(2)$ gauge theory $n_e$ \cite{ahk}, by
\begin{eqnarray}
n_e = p/2 \,.
\end{eqnarray}
Note that $n_e$ is properly thought of as being a linear combination
of the $U(1)$ factors on each 5-brane, which is why it is invariant as
the ends of the junction are moved around the inside of the face.

The $E_n$ weight vector $\lambda$ will characterize the (finite)
global symmetry quantum numbers of the state.  We are still left with
the grade $m$, which we now claim is irrelevant to the 5D theory.

Consider a junction with $p$ $(1,0)$ prongs on the top D5-brane.  Take
any one of these prongs and slide it along the loop of 5-brane until
it has gone once around and returns to the top.  During this process,
it must pass through all the branch cuts, and thus its invariant
charges change exactly by the addition of the charges of $\mdelta$; in
other words $m$ changes by 1 with other quantities fixed.  However, we
do not expect any of the field theory quantum numbers to be changed
under this process.  Hence when a junction ends on a 5-brane loop, the
quantity $m$ ceases to be an invariant, but instead can be changed
through continuous transformations.  The final value of $m$ will be
determined simply by the junction minimizing its length.  Hence this
parameter does not exist in the 5D field theory, consistent with the
fact that it has not been observed by other methods.  Note how in the
junction intersection form (\ref{affineinter}), all dependence on $m$
drops out when $k=0$, and it reduces to the same intersection form as
the finite case (\ref{intersection}).

A D3-brane in the vicinity of an $E_n$ 7-brane configuration will
realize a four-dimensional ${\cal N} = 2$ theory with $E_n$ global
symmetry.  The junctions leaving the 7-branes with $q=0$ are in
one-to-one correspondence with the ones realizing the BPS spectrum
from $\wE_n$ configurations in the five-dimensional case; thus the
electric spectrum in the four-dimensional theory is the same as that
in the five-dimensional theory.  This is as we would expect from the
del Pezzo picture, since both come from wrapping 2-branes on 2-cycles
in the del Pezzo, either in M-Theory or Type IIA string theory.

\section{E1 states from junctions}

Let us see in detail how the BPS states of the $E_1$ theory are
realized as junctions.

The basic states are the W-boson and the instanton, which we converted
into 5-7 junctions in \figref{5-d-5}.  We can read off the invariant
charges $Q^i$ and obtain the quantum numbers using the techniques of
\cite{dewolfezwiebach, dewolfe}:
\begin{eqnarray}
p &=& Q_{B1} + Q_{B2} + Q_{C1} + Q_{C2} \,, \\
a &=& Q_{C1} - Q_{C2} + 2 Q_{B2} \,. \nn
\end{eqnarray}
We find 
\begin{eqnarray}
\mbox{W-boson:} \quad \quad p=2& , \quad a = 1 \,, \\
\mbox{Instanton:} \quad \quad p=2& , \quad a=-1 \,, \nn
\end{eqnarray}
where $a$ is the single $E_1 = su(2)$ Dynkin label characterizing
$\lambda$.  We see both have $n_e=1$, and that they fall into a
doublet of the global symmetry, exactly as we expect.  Both junctions
have $\mJ^2 = 0$, with $g=0$ and $b=2$.  The instanton number used in
\cite{ahk} is given by $I = \half (p/2 - a)$.

Other BPS states are linear combinations of these two.  Under 
such combinations, the invariant charges add, and since quantities 
$p$ and $a$ are linear in the invariant charges, they will also
add.  Hence the charges and global symmetry representations
of arbitrary BPS states can be deduced easily from the junction
perspective.

Which junctions $\mJ$ are permitted as BPS states?  The most important
constraint is (\ref{jsquared}), which becomes
\begin{eqnarray}
(\mJ, \mJ) = - \fracs12 a^2 + \fracs18 p^2 \geq -2 + p \,.
\end{eqnarray}
In terms of the five-dimensional parameters $n_e$ and $I$, this is
\begin{eqnarray}
\label{e1self}
n_e (I-1) \geq (I+1) (I-1) \,,
\end{eqnarray}
and the set of permitted states precisely coincides with those 
given by \cite{kolrahmfeld}.  In addition, here we have shown
definitively that bound states of $m$ W-bosons, $m > 1$, with
no instantons are not BPS, without having to appeal to reducibility.

It should not be too surprising that the results coincide.  The
spectrum was determined in \cite{kolrahmfeld} using the intersection
numbers of curves in del Pezzo surfaces, which we will describe
further shortly.  Our results come ultimately from intersection
numbers determined by curves in K3.  Once the basis curves are
identified in each geometrical setup, and their intersection numbers
shown to agree, the inner product on both homology lattices agrees and
so the constraints should coincide.  However, \cite{kolrahmfeld} used
a mutual intersection condition, while we employ a self-intersection
condition.  In other cases (such as $\tilde{E}_1$) we must supplement
the self-intersection constraint with a mutual intersection
constraint.  This is described in the next section.

\section{Del Pezzo surfaces and BPS spectra}
\label{bps}

The five-dimensional theories we have been studying can be obtained by
compactifying M-Theory on a Calabi-Yau threefold containing a
shrinking four-cycle called a del Pezzo surface.  The $E_0$ theory is
obtained when the del Pezzo is just $\bbbp^2$, and the $\tilde{E}_1$,
$E_{2 \leq n \leq 8}$ theories correspond to $\bbbp^2$ blown up at $n$
generic points, denoted ${\cal B}_n$.  Finally, for $E_1$ the
four-cycle is $\bbbp^1 \times \bbbp^1$.  The BPS particle states in
the field theory are obtained by wrapping M2-branes on various
2-cycles within the del Pezzo.  Hence the second homology lattice of
del Pezzo surfaces determines the BPS spectrum.  The intersection of
these curves coincides with the intersection of the junctions
associated to the same BPS states.

An $n+1$-dimensional basis for the 2-cycles in ${\cal B}_n$ is
$\{\ell,e_1,..,e_n\}$ where $\ell$ is a $\bbbp^1$ inside $\bbbp^2$ and
the $e_i$ are the exceptional divisors of the blown-up points.  The
intersection matrix in this basis is diag $(1,-1,...,-1)$ and the
canonical class is given by $K_{{\cal B}_n} =
-3\ell+\sum_{i=1}^{n}e_{i}$.  For $\bbbp^1 \times \bbbp^1$ the basis
is $\ell_1$ and $\ell_2$, with $\ell_1 \cdot \ell_1 = \ell_2 \cdot
\ell_2 = 0$, $\ell_1 \cdot \ell_2$ = 1, and canonical class $K_{P^1
\times P^1} = -2 (\ell_1 + \ell_2)$.  We denote a general del Pezzo by
$X$.

The degree $d_C$ of a curve $C$ is defined by intersection with the
canonical class $K_X$,
\begin{eqnarray}
d_C = - C \cdot K_X \,.
\end{eqnarray} 
We will identify $d$ with the electric charge of the corresponding
state, $d = p = 2 n_e$.

In each case the homology lattice contains the root lattice of the
corresponding $E_n$ algebra. The set of roots is defined as curves
with vanishing degree and self-intersection $-2$,
\be
\label{delpezzoroots}
\{\alpha_i\} = \{C\in H^{1,1}(X) | (C\cdot C)=-2, (C\cdot K_X)=0\} \,,
\ee
and as such they generate a degree-preserving $E_n$ Weyl group action
on all curves.

Explicitly the simple roots for $E_{n}$, $\{3\leq n\leq 8\}$ are 
\be
\alpha_i=e_i-e_{i+1}\,, \quad i=1, \ldots ,n-1 \quad \mbox{and} \quad
\alpha_n = \ell - e_1-e_2-e_3 \,.
\ee
For $E_2$ the $su(2)$ root is $e_1 - e_2$ and for $E_1$ it is $\ell_1 - 
\ell_2$.
The Dynkin labels $a_i$ of a curve characterizing its $E_n$ weight
$\lambda$ are given by
\be
a_i=-C\cdot \alpha_i\,, \; i=1,\ldots,n \,.
\ee
In the $E_2$ and $\tilde{E}_1$ cases the $u(1)$ factor is not
associated with a root, but a curve not satisfying
(\ref{delpezzoroots}) can be chosen to give the appropriate
generalized Dynkin label. An arbitrary curve $C$ thus can be specified
by its $n$ Dynkin labels and its degree $d$, in terms of which the
intersection of two curves $C$ and $C'$ is
\be
\label{delpezzointer}
(C\cdot C')=-\lambda\cdot \lambda' + \frac{d \, d'}{9-n}.
\ee

With the identification $p \leftrightarrow d$, this coincides exactly
with the inner product on junctions given in (\ref{affineinter}) with
$f$ given in (\ref{f}) and $k=-q=0$.  Thus with each BPS state in an
$E_n$ five-dimensional theory, we can associate both a junction on a
$\wE_n$ set of 7-branes and a holomorphic curve in the corresponding
del Pezzo surface.

\subsection{Mutual intersection constraint}

Unlike in $K3$, in a del Pezzo surface $X$ self-intersection number
alone is not sufficient to determine if a given homology class has a
holomorphic representative or not. Holomorphic curves must satisfy the
adjunction formula:
\be
\label{adjunction}
C\cdot C+C\cdot K_{X}=2g-2 \,, \ee which corresponds to
(\ref{jsquared}), with $d$ substituted for gcd$(p,0) = p$.  Since K3
is Calabi-Yau and its canonical class vanishes, a corresponding
expression (\ref{adjunction}) would have no term $C \cdot K$, but it
is compensated for exactly by the inclusion of curves with boundary
components.

In addition, del Pezzo surfaces have a mutual intersection constraint:
holomorphic curves have positive degree and are required to have
positive intersection with all curves of degree one and
self-intersection number minus one \cite{mnw,hartshorne}.  (Curves of
negative degree can be antiholomorphic, and are just the negative of
the holomorphic curves.)  We will now derive a single constraint
ensuring this requirement, simplifying the multiple constraints of
\cite{mnw}.

Curves with $C \cdot C = -1$ satisfying (\ref{adjunction}) with
positive degree are necessarily $g=0$ and $d=1$. By
(\ref{delpezzointer}), they fall into a representation $R(n)$ of $E_n$
with
\begin{eqnarray}
\lambda^{2}=\frac{10-n}{9-n} \,.
\end{eqnarray}
For $E_{3\leq n \leq 8}$ these representations are ${\bf (3,2), 10,
16, 27, 56, 248}$.  They can be thought of as the fundamental
representations of $E_n$.  No such curves exist for $E_0$ and $E_1$
theories, so the mutual intersection constraint is trivial.  For
$\tilde{E}_1$ and $E_2$ there are $U(1)$ factors rendering the
definition of $\lambda$ conventional; we shall consider these theories
individually.

In general it is tedious to check the positivity of the intersection
of some curve $C = (\lambda, d)$ with every curve in $R(n)$.  However,
we can simplify the task as follows.  We wish to ascertain
\be
\label{condition}
(C,C_{i})=-\lambda\cdot \lambda_{i}+ \frac{d}{9-n}\geq 0
\ee
for all $\lambda_i \in R(n)$, or in other words,
\be
d\geq (9-n)\lambda\cdot \lambda_{i} \,.
\ee
Since the multiplication of weights is just the Euclidean inner
product, the most stringent condition arises when $\lambda$ and
$\lambda_i$ are in the same Weyl chamber.  By a Weyl rotation we can
make this the fundamental Weyl chamber, and thereby ascertain
(\ref{condition}) for all $\lambda$ in a given representation $R$ by
checking the intersection of the highest weight $\lambda_h(R)$ with just
the highest weight of $R(n)$:
\be
d\geq (9-n)\, \lambda_{h}(R)\cdot \lambda_{h}(R(n)) \,.
\ee

One must be careful to notice that for the curves with $g=0$, $d=1$
themselves, the mutual intersection constraint becomes
self-intersection, which is less stringent.  For example for the case
$n=8$, $R ={\bf 248}$ (\ref{condition}) seems to imply we need $d \geq
2$, but in fact since $R=R(8)$ itself this is not a mutual
intersection constraint, and we can have $d \geq 1$.

We can translate this directly into a mutual-intersection constraint
to be imposed on string junctions $\mJ = (\lambda, p)$ in the
$(p,q)$ 5-brane/7-brane setup:
\begin{eqnarray}
\label{mutual}
p\geq \lambda_{h}(R)\cdot \lambda_{h}(R(n)) \,.
\end{eqnarray}

\subsection{Relating D7-branes to geometrical blow-ups}
In this section we will relate the geometric picture with the brane
picture 
by establishing a correspondence between blowing up points 
and the addition of D7-branes. 

\onefigure{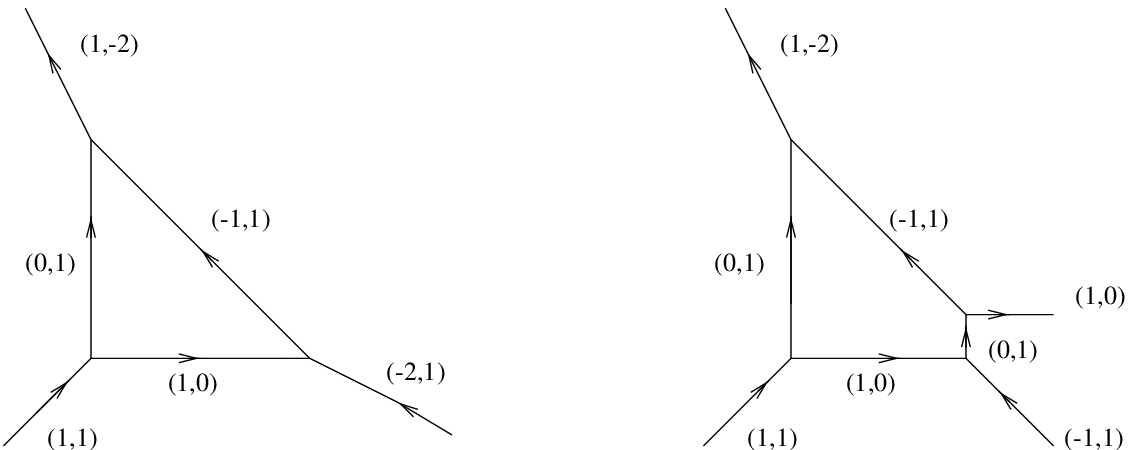}{a.\ The toric skeleton for $\bbbp^2$, which gives the
$E_0$ theory; the slopes are the degenerating cycles.  This is
identical to the corresponding $(p,q)$ 5-brane web. b.\ The toric
skeleton for ${\cal B}_1$, $\bbbp^2$ blown up at a generic point, here
chosen to be on the boundary.}
The del Pezzos which admit a description in toric geometry are the
ones associated to $E_0$, $\tilde{E}_1$, $E_1$, $E_2$ and $E_3$,
precisely those theories which can be realized as a $(p,q)$ 5-brane
web without 7-branes.  Consider $\bbbp^{2}$ and its one-dimensional
homology lattice generated by $\ell$ with $\ell^{2}=1$.  This
manifold admits a toric description, with a toric skeleton looking
exactly like the 5-brane web, \figref{5-d-1}a.  We can blow up a
generic point of $\bbbp^2$ and obtain the ${\cal B}_1$ del Pezzo
surface, introducing the new 2-cycle $e$, with $e^2 = -1$, $\ell \cdot
e = 0$, as \figref{5-d-1}b.

Notice to obtain ${\cal B}_1$ we blew up a point on the boundary of
the face.  This is not a special point on $\bbbp^2$, only in our toric
description of it.  It is not possible to represent torically a blown
up generic point inside the face; only the corners of the toric
diagram of $\bbbp^2$ can be blown up by replacing the point with a
$\bbbp^1$ which would be represented by a line segment in the toric
diagram. But due to an $SL(3,\bbbc)$ symmetry of $\bbbp^2$ any point
can be mapped to the point on the corner, and thus a generic blow-up
of $\bbbp^2$ can be represented by a toric diagram after this
$SL(3,\bbbc)$ transformation. However, this process cannot be
continued indefinitely since the $SL(3,\bbbc)$ symmetry is exhausted
after three blow-ups. Therefore there is no toric description of
$\bbbp^2$ blown up at more than three points; this creates the
familiar parallel external legs in the toric skeleton, which in a
toric description means the manifold no longer can go to zero size
inside the Calabi-Yau \cite{leungvafa}.

Adding 7-branes to the 5-brane web, we can avoid this problem,
effectively blowing up points on the interior of the face and
constructing all the $E_n$ theories.

\onefigure{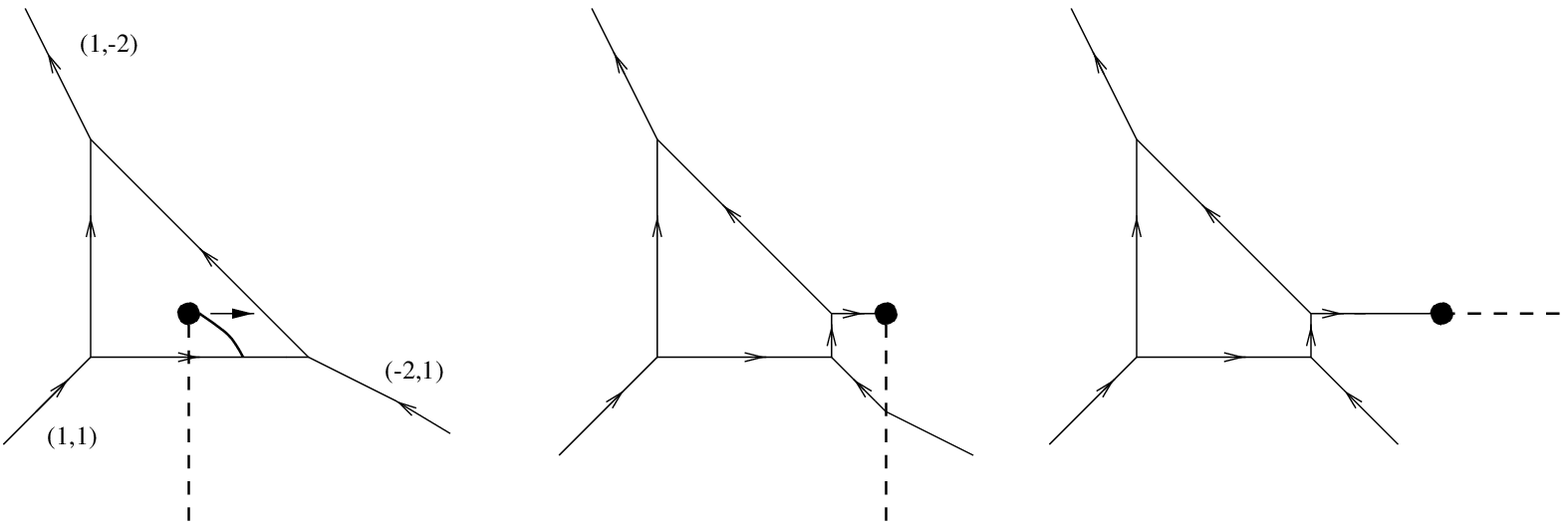}{Adding a D7-brane to the $(p,q)$ 5-brane web of
$E_0$ corresponds to blowing it up to $\tilde{E}_1$ at an interior
point, with a new flavor appearing as a string ending on the D7;
moving the D7-brane outside the face returns us to the usual blow-up
at the edge.}

We claim that adding a D7-brane in the interior of the 5-brane web
corresponds to blowing up a generic point on the del Pezzo.
Geometrically, a blow-up introduces a genus zero homology cycle of
self-intersection minus one.  Correspondingly, adding a D7-brane
introduces a new dimension to the junction lattice, generated by the
string leaving the D7-brane, which indeed has genus zero and
self-intersection $-1$.  Moving this D7-brane outside the 5-brane web
generates an external D5-brane (\figref{5-d-2}) and returns to the
picture with the blow-up on the boundary of the toric diagram.
However, we are now free to add more D7-branes beyond the point where
we are able to bring them outside the face, corresponding to blowing
up points beyond ${\cal B}_3$.  We can continue to add D7-branes
inside the face all the way to $E_8$, thus solving the problem of a
brane description of these theories.

A more precise geometrical interpretation of a general $(p,q)$ 7-brane
in the toric picture is still unknown, and would be very interesting
to determine.  The obstruction to proceeding beyond $E_8$ will be
explored in section \ref{fixedpoint}.

\subsection{BPS spectra for $E_n$ theories}

Here we summarize the BPS spectra for all the five-dimensional $E_n$
theories.  For $E_0$, $E_1$ and $\tilde{E}_1$ this has already been
discussed in \cite{kolrahmfeld}.  Related work on $E_6$, $E_7$ and
$E_8$ spectra can be found in \cite{mnw}.  We will review the results
for the first few theories, and then concentrate on $E_{n \geq 2}$.

As discussed previously, there is no additional mutual intersection
constraint for $E_0$ and $E_1$, and the self-intersection constraint
(\ref{jsquared}) suffices.  The basic BPS state of the $E_0$ theory is
the state $\Delta$ with $n_e = 3/2$, corresponding to the cycle
$\ell$.  All states $m \, \Delta$ for integer $m$ are permitted by
(\ref{jsquared}).  The lack of a W-boson demonstrates this is not a
gauge theory.

$E_1$ is pure $SU(2)$ with $\theta = 0$, and the del Pezzo is $\bbbp^1
\times \bbbp^1$.  The basic states are the W-boson and the instanton,
corresponding to the two curves $\ell_1$ and $\ell_2$, each with $n_e
=1$ and the latter with unit instanton number $I$.  The
self-intersection constraint evaluates to (\ref{e1self}), permitting
any state $n \ell_1 + m \ell_2$ with $n$, $m$ the same sign, except
when either integer vanishes, in which case the other must be $\pm 1$.

$\tilde{E}_1$ is the pure $SU(2)$ theory with $\theta = \pi$, but the
del Pezzo is that of the $E_0$ theory blown up at a point.  $\Delta$
appears as the curve $\ell$, and the curve $e$ arises, the associated
state being the ``instantonic quark'' $I_Q$ with $n_e = 1$; the
W-boson is the linear combination $\ell - e$.  One characterizes the
states by $n_e$ and the instanton number $I$, with $I=0$ for the
W-boson and $I=1$ for $\Delta$, $I_Q$.  The mutual intersection
constraint with the single curve $e$, combined with the
self-intersection constraint, forces all BPS states to be of the form
$n \ell + m (\ell - e)$ with $n$, $m$ of the same sign, with the
exception of $e$ itself, which is also BPS, and the states with $n=0$,
which must have $m=\pm 1$.

In \cite{kolrahmfeld}, the identical $\tilde{E}_1$ spectrum was
obtained by imposing two mutual intersection constraints, positivity
of intersection with $e$ and in addition with $\ell - e$, and no
self-intersection constraint; the states $n=0$, $|m| > 1$ were then
argued not to be BPS states since they are reducible.  It is
gratifying that we can reproduce these results with our methods, which
apply equally well to every $E_n$ theory, and which need not be
supplemented by appeals to reducibility.

States in the $E_{n \geq 2}$ theories are characterized by electric
charge $n_e$, instanton number $I$, and $n-1$ global symmetry charges
associated to quarks, $Q_{f_i}$.  These should correspond to
junction/del Pezzo quantities.  We have stated that $p = n_e/2$; the
instanton number $I$ and the $Q_{f_i}$ will be determined by the
weight vector $\lambda$.

Specifically, it is easy to see from \figref{5-d-2} that the quarks
all end on the D7-brane that adds the flavor.  Thus, the quark charge
$Q_{f_i}$ is just the invariant charge $Q_{Ai}$ on the appropriate
D7-(or $\mA$-)brane:
\begin{eqnarray}
Q_{f_i} = Q_{Ai} \,.
\end{eqnarray}
For $E_1$ we had $I = \fracs12(p/2 - a)$ where $a$ was the $SU(2)$
Dynkin label.  We can think of all junctions in $E_{n \geq 2}$
theories as the sum of an $E_1$ junction $\mJ_{E_1}$ with some $p'$
and $a$, thus determining $I$, and various strings from the
$\mA$-branes, determining $Q_{f_i}$: the total charge is then $n_e =
p/2 = (p' + \sum_i \, Q_{f_i})/2$.  The total self-intersection is
given by (\ref{intersection1})
\begin{eqnarray}
\mJ^2 = -\fracs12 a^2 + \fracs18 (p')^2 - \sum_i \, (Q_{f_i})^2 \,,
\end{eqnarray}
where the determinants vanish since the asymptotic strings are
mutually local.  Then in terms of $n_e$, $I$ and $Q_{f_i}$ the
self-intersection constraint (\ref{jsquared}) is
\begin{eqnarray}
I (2n_e - 2I - \sum_i Q_{f_i}) - \sum_i (Q_{f_i})^2 \geq 2 (n_e - 1)
\,,
\end{eqnarray}
or
\begin{eqnarray}
2 n_e (I-1) \geq 2 (I+1) (I-1) + I \sum_i Q_{f_i} + \sum_i (Q_{f_i})^2
\,.
\end{eqnarray}

The mutual intersection constraint is most easily evaluated for an
entire representation.  To this end, we provide the explicit map
between $I$, $Q_{f_i}$ and $\lambda$, and then list the constraint on
the weight vector.

For $E_{n \geq 3}$, we have
\begin{eqnarray}
\label{dynkins}
a_1 &=& n_e - 2I - \fracs12 \sum_i Q_{f_i} \,, \\ \nn
a_2 &=& Q_{f_1} + Q_{f_2} + I \,, \\
a_k &=& Q_{f_{k-1}} - Q_{f_{k-2}} \,, \quad k = 3 \ldots n \,. \nn
\end{eqnarray}

For all these cases, the mutual intersection constraint (\ref{mutual})
becomes
\begin{eqnarray}
p \geq 2 a_1 + 4 a_2 + 3 a_3 + 6 a_4 + 5 a_5 + 4 a_6 + 3 a_7 + 2 a_8 \,,
\end{eqnarray}
where for $n < 8$ one merely ignores the terms with $a_i$, $i > n$.
These numbers are exactly the Coxeter labels of $E_8$; for the other
$E_n$ the truncated set is the highest weight of $R(n)$ expanded in
the basis of simple roots.  Plugging in (\ref{dynkins}), we obtain
that the highest weight must satisfy
\begin{eqnarray}
\label{enQ}
Q_{f_{n-1}} \leq 0 \,.
\end{eqnarray}

The result (\ref{enQ}) is transparent from the junction point of view,
using (\ref{intersection1}) and the fact that the highest weight state
of $R(n)$ is simply the open string beginning on the last D7-brane.

For $E_2$, we have
\begin{eqnarray}
\label{e2dynkins}
a_1 &=& n_e - 2I - \fracs12 Q_f \,, \\
q &=& 2 Q_f + I \,, \nn
\end{eqnarray}
where $q$ is the $U(1)$ charge.  However the $U(1)$ factor destroys
our ability to classify the states by the Weyl orbits of a root
system.  Instead we give the allowed states explicitly.  The basis
states are the W-boson $\ell_1$ with $n_e = 1$, the instanton $\ell_2$
with $n_e = 1$ and $I=1$, and the quark $e$ with $n_e = \fracs12$ and
$Q_f = 1$.

The self-intersection constraint is
\begin{eqnarray}
2(n_e - I - Q_f/2)(I-1) \geq Q_f (Q_f + 1) \,,
\end{eqnarray}
and mutual intersection constrains $I>0$, $Q_f < 0$, and $2n_e \geq 2I
+ Q_f$; the negative of any state is also BPS.

\section{Approaching the fixed point and global symmetry}
\label{fixedpoint}

Thus far, we have described how one may use 7-branes to learn about
the global symmetry representations of the 5D states.  However, we
have not commented on how these symmetries are realized in the 5D
theory.  We recall that symmetry enhancement occurs only when 7-branes
are coincident; this is reminiscent of the situation in
\cite{hananywitten}, where symmetry enhancement was only explicit from
the string point of view when the D5-branes were coincident.  In the
K3 language, coincident branes correspond to colliding degenerate
fibers forming a singularity.  The types of possible deformable
singularities where classified by Kodaira.  Since it is clear that
affine groups do not appear in Kodaira's classification, none of the
7-brane configurations we have considered are fully collapsible
\cite{infinite}.  On the other hand, at the conformal fixed point
theory no distance scales should be present.

\onefigure{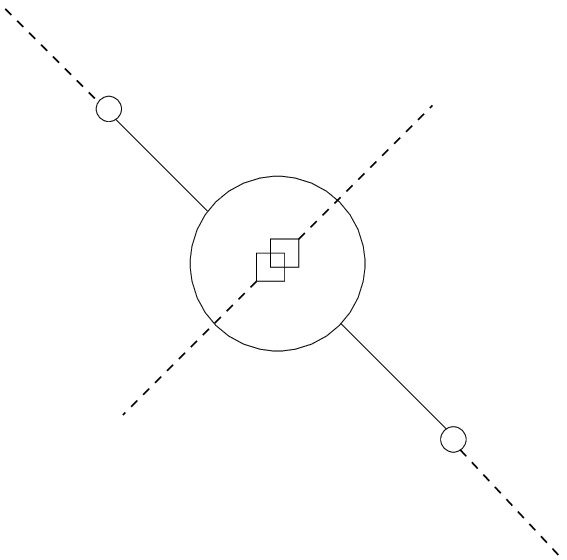}{As the $E_1$ system approaches the fixed point, only
two 7-branes can collapse, restoring the global symmetry.}  To solve
this puzzle let us return to the $E_1$ configuration of
\figref{5-d-4}.  The 7-branes may move along certain geodesics,
corresponding to the path that was followed by the external 5-brane
legs as modified by the nontrivial metric, without affecting the 5D
theory.  Hence, we may attempt to put the 7-branes on top of one
another by sliding them along these geodesics.  By inspecting the
configuration in \figref{5-d-4}, it is clear that 7-branes may
encounter each other only at infinite coupling, when the parameters
are tuned such that the face is a square.  We may choose the two
$\mC$-branes to be coincident, and ask what happens to the remaining
$\mB$-branes.

Since $\mB\mC\mC$ is not collapsible, once the two $\mC$-branes are
placed in middle of the face, the other branes must always remain a
finite distance away.  As the modulus is decreased, with the face
closing towards the fixed point, 5-brane prongs will inevitably form
on the $\mB$-branes (\figref{5-d-6}).  The BPS states of the 5D theory
will either be 5-5 strings or 5-7 strings emanating from the
$\mC$-branes.  When the face closes, all these become massless.  On
the other hand, the $\mB$-branes seem superficially to provide a
distance scale, spoiling the conformal fixed point.  However, since
the theory (as before) is insensitive to the position of the
$\mB$-branes along their geodesics, the 5D theory is ignorant of this
scale.

This is a crucial point for product group symmetry, as in $E_3$ for
instance.  There, the global $SU(2) \times SU(3)$ is achieved by 2
branes of one type coinciding and 3 branes of another type coinciding
(happening only at infinite coupling and zero bare masses).  The two
types are mutually non-local and non-collapsible.  Therefore, it is
not possible for both sets of coincident branes to be in the center of
the face.  Instead, the face will close in between the two sets, with
external 5-brane legs on the two sets of branes.  Again, the 5D
theory's ignorance of the distance along the external legs ensures
that the symmetry is in fact the full product, rather than just one of
its components. Because the legs may be extended to infinity, the
fixed point picture is essentially the original 5-brane web without
7-branes.  More generally, since there are no product groups in
Kodaira's classification, we see that a product group fixed point is
possible, only if there is a 5-brane web without 7-branes to realize
it.

As discussed, the higher $E_n$ series ($n \geq 4$) have no description
solely in terms of 5-branes, and as a result as the theory approaches
the origin of the Coulomb branch we do not expect the 5-7 brane system
to revert to the 5-brane web without 7-branes.  Instead, the global
symmetry must originate from a loop closing on a fully collapsible
portion of the original affine configuration.  In particular, this
implies that there should not be product groups among the higher $E_n$
series, which is the case.  Indeed, all $n \geq 4$ configurations may
be represented by a single set of collapsible branes responsible for
the symmetry, and a few remaining branes; as this is somewhat
technical we present it in an appendix.  These remaining branes,
although essential to allowing a closed 5-brane loop (and thus a 5D
theory) to exist, play no role in the symmetry.  For these cases, the
fixed point picture has the face closing upon a single point - the
location of the global symmetry branes.  The remaining branes will be
connected by external legs to the 5-brane loop.  This picture makes it
manifest that the states becoming massless indeed transform under the
expected global symmetry.

\onefigure{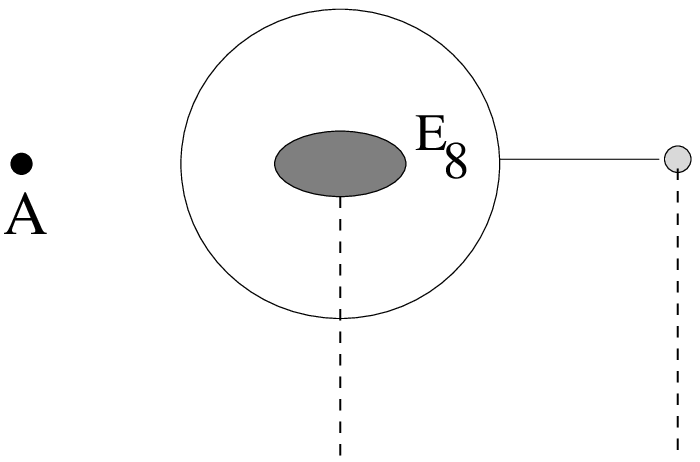}{A configuration with 8 flavors is not associated
with a new fixed point, as an $E_9$ configuration of 7-branes cannot
collapse, and a scale associated to a D7-brane (A) remains.}  In
addition, one may use the above picture to understand why the maximum
number of flavors is 7.  Adding flavors is achieved by adding
$\mA$-branes to the $E_1$ configuration.  First, we must ask if there
are restrictions in constructing a 7-brane background with any number
of $\mA$-branes.  Asymptotically, the monodromy around an affine
configuration prescribes the complex parameter to be
\begin{equation}
\tau(z) = \frac{i}{2\pi}(8-n)\log(z) + \tau_0.
\end{equation}
Here, $n$ is the number of $\mA$-branes, $|z|>1$, and $\tau_0$ is a
constant.  The metric around the 7-branes is proportional to Im$\tau$,
which therefore must always be positive.  This requires $n\leq 8$.

The case of $n=8$ is more subtle.  This is the so-called $\wE_9$
configuration, realizing a doubly-affinized $E_8$ algebra
\cite{dewolfe,infinite}, with trivial monodromy and the ability to
support a string loop of any charge winding around the configuration.
Only the $E_8$ part is collapsible; in addition there is the
affinizing brane that made $\wE_8$ and the $\mA$-brane associated to
the eighth flavor that made $\wE_9$, and these two branes stay a
finite distance from the $E_8$.  As the face shrinks upon the
collapsed $E_8$ point, the loop will form an external leg on the
affinizing brane as in the 7-flavor ($E_8$) case, but will completely
ignore the $\mA$-brane, since they are mutually local.  Thus, the
distance of the $\mA$-brane, corresponding to the mass parameter of
the eighth flavor, stays as a scale in the 5D theory and can never be
set to zero.  One can only reach a conformal fixed point by integrating
out this flavor, thus returning to the $E_8$ case.

\subsection*{Acknowledgements} 

We are grateful for conversations with Tam\'as Hauer, Andreas Karch,
Barak Kol, Lisa Randall, Cumrun Vafa, and Barton Zwiebach. This work
was supported by the U.S.\ Department of Energy under contract
\#DE-FC02-94ER40818.

\newpage
\section*{Appendix: 7-brane Technology}
In this section we will review some of the properties of the 7-brane 
configurations realizing the $\widehat{E}_{n}$ symmetry. We will show that
it is always possible to find a sub-configuration of 7-branes realizing 
$E_{n}$ symmetry which can be collapsed. 

%{\bf Monodromy:}
A $[p,q]$ 7-brane has an $\sl2z$ monodromy matrix given by
\be
K_{[p,q]}=\pmatrix{1+pq & -p^{2}\cr q^{2} &1-pq}.
\ee
An $\ch{r}{s}$-string crossing the branch cut of the $[p,q]$ 7-brane is 
converted into an $K_{[p,q]} \ch{r}{s}$-string.

The canonical presentation of 7-brane configurations is the one in
which the branes are located along the real axis with branch cuts
going downwards.  This is convenient for discussing symmetry
properties.  With this convention the total monodromy around the brane
configuration \be \mX_{\bf [p_{1},q_{1}]} \ldots \mX_{\bf
[p_{n-1},q_{n-1}]} \mX_{\bf [p_{n},q_{n}]} \,, \ee

is given by \be
K=K_{[p_{n},q_{n}]}K_{[p_{n-1},q_{n-1}]}\ldots K_{[p_{1},q_{1}]}.  \ee

%{\bf Equivalence:}
Because of the $SL(2,\bbbz)$ symmetry of Type IIB
string theory and the arbitrariness of the location of branch cuts,
two configurations of 7-branes are {\em equivalent} if they can be
transformed into each other via $SL(2,\bbbz)$ transformations and
branch cut moves.  Equivalent configurations naturally realize the
same algebra.

{\bf Global transformation:} The monodromy matrix $K$ of a 7-brane can
also be expressed in a charge-vector notation.  If $\mz^T \equiv
[p,q]$ is the 7-brane charges, \be K_{\bf z}=1+\mz^{T}\mz \, S,
\quad \mbox{where} \quad S=\pmatrix{0 & -1\cr 1&0} \,.
\label{monodromy}
\ee Under a global transformation $g \in SL(2,\bbbz)$, we have $\mz
\ra g \, \mz$.  It is easy to see from (\ref{monodromy}) that the
monodromy undergoes a conjugation by $g$, \be K_{g{\bf z}}=gK_{\bf
z}g^{-1} \ee

{\bf Branch cut moves:} The labeling of branes actually depends on the
placement of branch cuts.  We can move the branch cut of one 7-brane
$\mX_{\zz_1}$ across another 7-brane $\mX_{\zz_2}$, thus changing the
latter to $\mX_{\zz'_2}$ and exchanging their order in the canonical
presentation, as explained in \cite{mgthbz}, and in the canonical
presentation they become
\begin{eqnarray}
\label{crosstrans} 
{\bf X}_{\displaystyle\zz_1}  {\bf X}_{ \displaystyle\zz_2}  &=& 
{\bf X}_{\displaystyle\zz_2 }\, 
{\bf X}_{\,\displaystyle (\zz_1 + (\zz_1 \times \zz_2)\,
\zz_2\, )} \nonumber \\
&=& {\bf X}_{ \,\displaystyle 
(\zz_2 + (\zz_1 \times \zz_2)\, \zz_1\, )}\, 
{\bf X}_{\displaystyle\zz_1 }\,,
\end{eqnarray}
where we have defined
\be
\label{introdet} 
{\bf z}_1 \mtimes {\bf z}_2 \equiv -{\bf z}_1^T S\, {\bf z}_2 = 
{\bf z}_2^T S\, {\bf z}_1 =
\hbox{det}\, \pmatrix{ p_1 & p_2 \cr q_1 & q_2}\,.  
\ee 
Equation (\ref{crosstrans}) indicates the fixed brane acquires an
extra charge equal to the charge of the moving brane times the
determinant of the relative charges.

\bigskip

%{\bf $E_{n}$ configurations:}
The 7-brane configurations associated
with $E_{n}$ symmetry will be denoted by ${\bf E_{n}}$. One useful
presentation of ${\bf E_{n}}$ is ${\bf A^{n-1}BCC}$ for $n>0$
\cite{gz}.  The intersection of junctions supported on ${\bf E_{n}}$
is given by (\ref{intersection}) with $f(p,q;p',q')$ given in
(\ref{f}).

For $n \geq 2$, the configuration ${\bf A^{n-1}BCC}$ is equivalent to
${\bf A^{n}X_{[2,-1]}C}$ \cite{finite}.  For $n=1$ the equivalence
does not hold; ${\bf E_1} = \mB\mC\mC$ and ${\bf \tilde{E}_{1}}={\bf
AX_{[2,-1]}C}$ are distinct configurations.  Furthermore we have ${\bf
E_{0}}={\bf X_{[2,-1]}C}$ which can be ``blown up'' only to ${\bf
\tilde{E}_1}$ by adding an $\mA$-brane.

%{\bf $\wE_n$ configurations:}
The ${\bf E_n}$ configuration can be
``affinized'' by adding another 7-brane to obtain $\wE_n$ symmetry:
\begin{eqnarray}
{\bf \wE_n} = \mA^{n-1} \mB\mC\mC \mX_{\bf [3,1]}= \mA^{n-1}
\mB\mC\mB\mC \,.
\end{eqnarray}
The affinizing brane can never be collapsed onto the others, so the
$\wE_n$ symmetry cannot be restored.

The affinizing brane is a spectator in the equivalence between the
two presentations of $E_n$, so we can analogously construct
${\bf \widehat{\tilde{E}}_1} = {\bf A^1X_{[2,-1]}CX_{[4,1]}}$ and
${\bf \wE_0} = {\bf X_{[2,-1]}CX_{[4,1]}} = {\bf \mX_{\bf [2,-1]}
\mX_{\bf [-1,2]} \mC}$.

We summarize the correspondence of these
brane configurations with the geometry of del Pezzos in the following
table.

\begin{center}
\begin{tabular}{||c|c|c||} \hline
5D Field Theory & Brane Configuration & Geometry\\ \hline\hline

$E_0$ & ${\bf E_0} = {\bf X_{[2,-1]}X_{[-1,2]} C}$
&$\bbbp^{2}\rule{0mm}{5mm}$ \\ \hline
$\tilde{E}_1$ &
${\bf AX_{[2,-1]}X_{[-1,2]}C}$ & ${\cal B}_{1}\rule{0mm}{5mm}$ \\ \hline

$E_1$ & ${\bf BCBC}$ &$\bbbp^{1}\times \bbbp^{1}\rule{0mm}{5mm}$\\ \hline

$E_n, n \geq 2$ & $\mA^{n-1}\mB\mC\mB\mC= \mA^{n} {\bf
X_{[2,-1]}CX_{[-1,2]}}$ & ${\cal B}_{n} \rule{0mm}{5mm}$ \\ \hline
\end{tabular}
\end{center}

Notice that while the 7-brane configurations for $E_n$ as given are
natural from the point of view of ``blowing-up'' by adding
$\mA$-branes as discussed in section \ref{bps}, some equivalent
configuration may be natural when the 7-branes are moved outside the
5-branes.

%{\bf Collapsible sub-configurations of $\widehat{E}_{n}$:}
In section
\ref{fixedpoint}, we stated that for each ${\bf \wE_n}$ there existed
a sub-configuration of collapsible 7-branes realizing the finite
symmetry $E_n$.  We are now in a position to describe these explicitly.
An equals sign denotes a branch cut move, while a $\cong$ with
an $SL(2,\bbbz)$ matrix above it denotes a global $SL(2,\bbbz)$.

For ${\bf \wE_{6}, \wE_{7}}$ and ${\bf \wE_{8}}$, the full finite
subconfigurations ${\bf E_6}$, ${\bf E_7}$, and ${\bf E_8}$ correspond
to Kodaira singularities and are collapsible.

For ${\bf \widehat{E}_{5}}$ we expect a collapsible 
sub-configuration realizing $E_{5}=D_{5}$.
The usual presentation of 7-branes realizing $D_{5}$ is 
${\bf A^{5}BC}={\bf D_{5}}$. We 
can see from the following branch cut moves and global $\sl2z$ that 
${\bf D_{5}}$ is indeed present in ${\bf \widehat{E}_{5}}$.
\bea
{\bf A^{4}BCBC}={\bf A^{5}X_{[2,-1]}CX_{[4,1]}}
\stackrel{T^{-2}}{\cong} {\bf A^{5}X_{[4,-1]}BX_{[2,1]}}
\\ ={\bf CA^{5}BX_{[2,1]}}={\bf (X_{[0,1]})^{5}CBX_{[2,1]}}
\stackrel{S}{\cong}{\bf A^{5}BCX_{[-1,2]}} \,. \nn
\eea
For ${\bf \widehat{E}_{4}}$ we expect a sub-configuration realizing
$A_{4}$.  The usual presentation of $A_{4}$ is ${\bf A^{5}}$. The
following branch cut moves and global $\sl2z$ makes it clear that such
a subconfiguration is indeed present in ${\bf \widehat{E}_{4}}$:
\bea
{\bf A^{3}BCBC}={\bf A^{4}X_{[2,-1]}CX_{[4,1]}}
&=&{\bf A^{4}X_{[2,-1]}X_{[-1,2]}C}
={\bf A^{4}X_{[5,-1]}X_{[2,-1]}C}= \nn \\{\bf BA^{4}X_{[2,-1]}C}
={\bf BX_{[2,-1]}B^{4}C}
&=&{\bf B^{5}X_{[-2,3]}C} \stackrel{g}{\cong}{\bf A^{5}X_{[-2,1]}X_{[1,2]}}.
\eea
with $g = \left({1\;0}\atop{1\;1}\right)$.
For ${\bf \widehat{E}_{3}}$ we have,
\bea
{\bf A^{2}BCBC}={\bf A^{3}X_{[2,-1]}CX_{[4,1]}}
={\bf A^{3}X_{[2,-1]}X_{[-1,2]}C}
&\stackrel{T}{\cong}& {\bf A^{3}BX_{[1,2]}X_{[2,1]}}= \nn \\{\bf B(X_{[0,1]})^{3}X_{[1,2]}X_{[2,1]}}
={\bf B^{2}(X_{[0,1]})^{3}X_{[2,1]}}
&\stackrel{g'}{\cong}& {\bf A^{2}(X_{[0,1]})^{3}X_{[2,3]}} \,,
\eea
with $g' = \left( {\;1\;\,0} \atop {\!-1\;1} \right)$.
{From} the above presentation of ${\bf \widehat{E}_{3}}$ we see that it has
a collapsible sub-configuration realizing $E_{3}=A_{1}\times A_{2}$.

For ${\bf \wE_1}$ and ${\bf \wE_2}$ there is a manifest $SU(2)$ when
two $\mC$-branes come together, as in section \ref{fixedpoint}.  The
$U(1)$ factors in ${\bf \wE_2}$ and ${\bf\widehat{\tilde{E}}_1}$ are
not associated with collapsing branes.

%%%%%%%%%%%%%%%%%%%%%%%%%%%%%%%%%%%%%%%%%%%%%%%%%%%%%%%%%%%%%%%%%%%%


\begin{thebibliography}{99}


\bibitem{seiberg} 
N.~Seiberg, {\it Five dimensional SUSY field
theories, non-trivial fixed points and string dynamics},
Phys.~Lett.~{\bf B388} (1996) 753, hep-th/9608111.

\bibitem{delpezzo}
M.R.~Douglas, S.~Katz and C.~Vafa, {\it Small instantons, Del Pezzo
surfaces and type I$'$ theory}, Nucl. Phys. {\bf B497}, 155 (1997),
hep-th/9609071.

\bibitem{morrisonseiberg}
D.R.~Morrison and N.~Seiberg, {\it Extremal transitions and
five-dimensional supersymmetric field theories}, Nucl. Phys. {\bf
B483}, 229 (1997), hep-th/9609070.

\bibitem{kmv}
A.~Klemm, P.~Mayr and C.~Vafa,
{\it BPS states of exceptional noncritical strings},
hep-th/9607139.

\bibitem{mnw}
J.A.~Minahan, D.~Nemeschansky and N.P.~Warner,
{\it Investigating the BPS spectrum of noncritical $E_n$ strings},
Nucl. Phys. {\bf B508}, 64 (1997), hep-th/9705237.


\bibitem{ah}
O.~Aharony and A.~Hanany, {\it Branes, superpotentials and
superconformal fixed points}, Nucl. Phys. {\bf B504}, 239 (1997),
hep-th/9704170.


\bibitem{ahk}
O. Aharony, A. Hanany, B. Kol, {\it Webs of $(p,q)$ 5-branes, Five
dimensional field theories and grid diagrams}, JHEP.{\bf 01} (1998)
002, hep-th/9710116;.

\bibitem{kol}
B.~Kol, {\it 5-D field theories and M theory,} hep-th/9705031.

\bibitem{kolrahmfeld}
B. Kol and J. Rahmfeld, {\it BPS spectrum of $5$ dimensional field
theories, $(p,q)$ webs and curve counting}, JHEP.{\bf 9808} (1998) 006,
hep-th/9801067.

\bibitem{leungvafa}
N.~C.~Leung and C.~Vafa, {\it Branes and toric geometry},
Adv.~Theor.~Math.~Phys.~{\bf 2} (1998) 91, hep-th/9711013.




\bibitem{gz} 
M.~R.~Gaberdiel, B.~Zwiebach, {\it Exceptional groups from open
strings}, Nucl.~Phys.~{\bf B518} (1998) 151, hep-th/9709013.

\bibitem{mgthbz}
M.~R.~Gaberdiel, T.~Hauer, B.~Zwiebach, {\it Open string-string
junction transitions}, Nucl.~Phys.~{\bf B525} (1998) 117,
hep-th/9801205.  


\bibitem{dewolfezwiebach}
O.~DeWolfe and B.~Zwiebach, {\it String junctions for arbitrary Lie
algebra representations} hep-th/9804210, 
to appear in Nucl.~Phys.~{\bf B}.  

\bibitem{dhiz} 
O.~DeWolfe, T.~Hauer, A.~Iqbal and B.~Zwiebach, {\it Constraints On
The BPS Spectrum Of N=2, D=4 Theories With A-D-E Flavor Symmetry},
hep-th/9805220, to appear in Nucl.~Phys.~{\bf B}.


\bibitem{dewolfe}
O.~DeWolfe, {\it Affine Lie Algebras, String Junctions And
7-Branes}, hep-th/9809026.

\bibitem{finite}
O.~DeWolfe, T.~Hauer, A.~Iqbal and B.~Zwiebach, {\it Uncovering the
symmetries on [p,q] seven-branes: Beyond the Kodaira classification},
hep-th/9812028.

\bibitem{infinite}
O.~DeWolfe, T.~Hauer, A.~Iqbal and B.~Zwiebach, {\it Uncovering infinite
symmetries on [p, q] 7-branes: Kac-Moody algebras and beyond},
hep-th/9812209.

\bibitem{nekrasov}
A.~Mikhailov, N.~Nekrasov, S.~Sethi, {\it
Geometric Realizations of BPS States in N=2 Theories}, hep-th/9803142.

\bibitem{ims}
K.~Intriligator, D.R.~Morrison and N.~Seiberg, {\it Five-dimensional
supersymmetric gauge theories and degenerations of Calabi-Yau
spaces}, Nucl. Phys. {\bf B497}, 56 (1997), hep-th/9702198.

\bibitem{telaviv}
A.~Brandhuber, N.~Itzhaki, J.~Sonnenschein, S.~Theisen, and
S.~Yankielowicz, {\it On the M theory approach to (compactified) 5-D
field theories}, Phys. Lett. {\bf B415}, 127 (1997), hep-th/9709010.


\bibitem{hananywitten}
A.~Hanany and E.~Witten, {\it Type IIB Superstrings, BPS Monopoles, and
three-dimensional gauge dynamics}, Nucl.~Phys.~{\bf B492} (1997) 152,
hep-th/9611230.

\bibitem{sen}
A.~Sen, {\it F-theory and Orientifolds}, Nucl.~Phys.~{\bf B475} (1996)
562-578, hep-th/9605150. 



\bibitem{minahan}
J.~A.~Minahan and D.~Nemeschansky,
{\it An N=2 superconformal fixed point with $E_6$ global symmetry},
Nucl.~Phys.~{\bf B482} (1996) 142, hep-th/9608047;
\hfill\break
J.~A.~Minahan and D.~Nemeschansky, {\it Superconformal fixed points
with $E_n$ global symmetry}, Nucl.~Phys.~{\bf B489} (1997) 24,
hep-th/9610076.


\bibitem{estrings}
J.~Minahan, D.~Nemeschansky, C.~Vafa and
N.~Warner, {\em Strings And N=4 Topological Yang-Mills Theories},
Nucl.~Phys.~{\bf B527} (1998) 581, hep-th/9802168.


\bibitem{hartshorne}
R.~Hartshorne, {\it Algebraic Geometry}, Springer-Verlag (1977).

\end{thebibliography}
\end{document}